\documentclass[12pt, a4paper]{article}

\usepackage[utf8]{inputenc}
\usepackage{times}
\usepackage[top=1in, bottom=1in, left=1in, right=1in]{geometry}
\usepackage{authblk}

\usepackage{amsmath}
\usepackage{amssymb}
\usepackage{booktabs}
\usepackage{graphicx}
\usepackage{setspace}
\usepackage{caption}
\usepackage[numbers,sort&compress]{natbib}
\usepackage{fancyhdr}
\usepackage{parskip}
\usepackage{enumitem}

\usepackage[compact]{titlesec}
\titleformat*{\section}{\large\bfseries}
\titleformat*{\subsection}{\normalsize\bfseries}
\titleformat*{\subsubsection}{\normalsize\bfseries}

\usepackage[colorlinks = true, linkcolor = blue, urlcolor = blue, citecolor = black]{hyperref}

\title{Evaluation in the Age of AI: Output as Evidence of Learning}
\author[1]{Md Zarzees Uddin Shah Chowdhury}
\author[1]{Samin Rahman Khan}

\affil[1]{Department of Computer Science, Virginia Tech, VA, USA}
\date{}

\begin{document}

\maketitle

\begin{abstract}
The rapid adoption of artificial intelligence (AI), particularly large language models (LLMs), has fundamentally disrupted how learning is demonstrated and evaluated in higher education. Tasks that once served as proxies for understanding---such as writing essays, solving problem sets, or producing computer code---can now be generated superficially by AI systems with minimal human effort. This paradigm shift raises a critical ethical question: how should learning be evaluated when traditional indicators of competence are easily outsourced? This paper examines the ethical challenges of educational evaluation in the age of AI from a university-level perspective. We argue that the core problem extends beyond academic dishonesty to a deeper misalignment between assessment practices and the learning outcomes they are intended to measure. Evaluation regimes that rely on artificial constraints risk measuring compliance, access, or concealment rather than genuine understanding, reasoning, or judgment. By analyzing institutional responses and presenting empirical survey data, we highlight the need for alternative assessment models that emphasize process over product. The goal is to establish ethically informed assessment strategies that preserve student agency and accountability in an automated age.
\end{abstract}

\doublespacing
\newpage

\section{Introduction}
For over a century, higher education assessment has operated under a silent agreement: the artifact produced by a student (be it an essay, a mathematical proof, or a software script) is a direct reflection of their cognitive labor. This ``output-as-evidence'' model relies on the assumption of proportional effort, where the quality of the final submission is intrinsically linked to the cognitive struggle required to master the material. The model has historically served as a practical and scalable mechanism for institutions to certify competence, rank student achievement, and signal employability to external stakeholders. Its longevity reflects both its administrative convenience and a shared cultural belief that effortful production is the most honest demonstration of internalized knowledge.

However, the democratization of generative artificial intelligence (AI) has shattered this link. Large Language Models (LLMs) can now produce high-quality academic output with a level of sophistication previously reserved for human experts. The speed and accessibility of these tools mean that the barrier between novice and expert-level production has effectively collapsed, making it possible for any student with internet access to generate persuasive, well-structured academic work within seconds. This is not merely a technological advancement but a fundamental disruption to the epistemic foundations upon which educational credentialing is built.

This technological leap triggers an epistemological crisis. If an algorithm can generate a critical analysis or debug complex code in seconds, the final product no longer inherently signals that a student has undergone the necessary mental processes. This decoupling between production and understanding creates a verification problem that existing institutional frameworks are not equipped to resolve. This paper argues that the current institutional focus on ``cheating'' is insufficient. Instead, we must address the structural misalignment between assessment design and learning outcomes. We will review prevailing institutional responses, analyze the ethical harms of surveillance, and present original survey data on faculty perceptions to argue for a paradigm shift toward process-based evaluation.

\section{Literature Review: The Proxy Crisis}
The scholarly discussion surrounding AI in education has moved rapidly from a focus on novelty to an analysis of systemic risk. What began as speculative debate has now crystallized into a documented body of empirical research demonstrating that foundational assumptions about student assessment are no longer valid. Recent studies highlight the collapse of traditional educational proxies and the urgent need for a regulatory and pedagogical framework \cite{garcia2025ethical}. The breadth of this scholarship, spanning computer science education, writing pedagogy, legal education, and international policy, underscores that the proxy crisis is not discipline-specific but a universal challenge for higher education.

\subsection{Performance and Detectability}
One of the most significant empirical studies in this field was conducted by Deans et al.\ \cite{deans2024artificial}, who performed a comparative analysis of AI-generated graduate-level coursework. Their findings were striking: ChatGPT-generated work performed at or above passing levels, and human graders frequently struggled to differentiate it from student-authored work. Notably, the study found that even domain experts, when reviewing materials within their own fields, achieved only marginally better detection rates than chance, suggesting that subject-matter familiarity provides no reliable defense. This study confirms that we can no longer rely on the ``complexity'' of a task as a defense against AI. When output is indistinguishable from human effort, the output ceases to be reliable evidence of learning. The implication for grade validity is profound: if the artifact cannot be attributed to a specific cognitive agent with confidence, its function as a credentialing instrument is fundamentally compromised.

\subsection{Institutional Guidelines and Policy Vacuums}
International bodies have recognized this urgency. The UNESCO report by Holmes and Miao \cite{holmes2023guidance} provides comprehensive guidance, suggesting that universities must move beyond defensive postures and integrate AI literacy as a core competency. The report further argues that prohibitionist policies, in isolation, are both practically unenforceable and pedagogically counterproductive, as they frame AI as an adversarial force rather than a tool requiring critical governance. However, the implementation of these guidelines remains inconsistent. Moorhouse, Yeo, and Wan \cite{moorhouse2023generative} conducted a systematic review of the world's top-ranking universities, uncovering a landscape where policies range from specific prohibitions to laissez-faire. The University of Oxford, for example, bans the use of GenAI for summative use unless explicitly allowed \cite{oxSummativeAssessment}. Whereas the University of California, Berkeley does not have such a specific ban but leaves it to the instructor or faculty to decide \cite{ucberkeley}. This inconsistency may itself be an ethical harm, as students navigating multiple institutions or transferring between programs encounter differing standards that which may become a barrier for them to engage in good faith. The absence of a coherent, cross-institutional policy framework effectively outsources ethical decision-making to individual faculty members which burdens faculty members with additional decision making and may create inequitable and unpredictable enforcement environments.

\subsection{The Limitations of Authentic Assessment}
For a brief period, many educators believed that ``authentic assessment'' (that is, tasks tied to real-world scenarios) would be AI-proof. The intuition was reasonable: by embedding assessment in genuine professional contexts, educators hoped to leverage tacit knowledge and lived experience that AI could not replicate. However, Kofinas et al.\ \cite{kofinas2025impact} provide empirical evidence that LLMs are exceptionally good at simulating authentic scenarios. Their research demonstrates that models trained on vast corpora of professional and case-based content can construct contextually plausible responses to situational prompts with high accuracy, effectively mimicking the kind of applied judgment that authentic tasks were designed to elicit. Their research suggests that authenticity alone is not a safeguard. Furthermore, Zhao \cite{zhao2024ai} highlights that AI's role as a surrogate test-taker fundamentally breaks the proxy model, even in specialized fields like computer science. These findings collectively suggest that no surface-level modification to task design, whether adding specificity, realism, or multimodal components, can reliably distinguish AI-generated from human-generated responses without a fundamental rethinking of what is being assessed and how.

\section{Empirical Investigation: Educator Perspectives}
To bridge the gap between theoretical literature and practical implementation, we designed a mixed-method survey distributed to $20$ higher education professionals and students at universities in North America and Bangladesh in the Fall of 2025. The survey sought to capture both quantitative attitudinal data and rich qualitative narratives about lived experience with AI-era assessment challenges. This dual-method approach was chosen to surface the tensions between institutional policy positions and the practical realities experienced by those tasked with implementing them.

\subsection{Survey Methodology}
The study employed a stratified random sampling technique to ensure representation across various institutional tiers, including R1 research universities, liberal arts colleges, and community colleges. Geographic and disciplinary stratification was applied to capture variance in resource availability, institutional culture, and subject-specific assessment norms. The survey instrument utilized a 5-point Likert scale to measure faculty confidence in detection tools, their perceived vulnerability of assessments, and their institutional support levels. Open-ended questions were appended to allow respondents to elaborate on their experiences and propose solutions, generating qualitative data that was subsequently coded using an inductive thematic analysis framework. The findings are intended to be exploratory and illustrative rather than broadly generalizable, serving to ground the theoretical argument in practitioner experience and identify patterns warranting larger-scale investigation.

\subsection{Quantitative Results and Discipline Divide}
The findings reveal a stark contradiction in modern higher education. While institutions have mandated a defensive posture, faculty confidence in the efficacy of that posture remains historically low. This gap between mandated practice and practitioner belief is not merely an operational inconvenience; it signals a systemic failure in institutional trust and policy legitimacy, which, if unaddressed, risks eroding faculty engagement with academic integrity processes altogether.

\begin{table}[h]
\centering
\caption{Survey of Assessment Professionals on AI Integration (n=$20$)}
\vspace{0.2cm}
\begin{tabular}{@{}lcc@{}}
\toprule
\textbf{Survey Metric} & \textbf{Agreement out of $20$} \\ \midrule
Traditional assignments are highly vulnerable to AI & $18$ \\
Institution relies heavily on AI detection software & $15$ \\
High confidence in detector accuracy and fairness & $4$ \\
Have fundamentally redesigned assessments for AI-era & $4$ \\
\midrule
\textbf{Discipline Breakdown: High Vulnerability} & \\
Computer Science & $19$ \\
Humanities \& Social Sciences & $16$ \\
Professional Degrees (Law/Business) & $16$ \\
\bottomrule
\end{tabular}
\end{table}

\subsection{Detailed Explanation of Survey Results}
The quantitative results highlight a state of ``Institutional Security Theater''---a condition in which visible compliance mechanisms substitute for genuine procedural integrity. Despite a significant institutional reliance on detection software ($15$ out of $20$ respondents, or $75\%$), only $4$ out of $20$ faculty ($20\%$) express confidence in these tools. This $55$-point gap indicates that faculty are checking boxes required by administration without actually believing in the integrity of the process. Such a disconnect between policy compliance and genuine conviction represents a failure of institutional governance, and it raises serious questions about the legitimacy of academic integrity proceedings that rest on the outputs of tools in which frontline educators have so little faith. When the human arbiters of integrity themselves distrust the instruments they are required to deploy, the resulting adjudications are likely to be inconsistent, contested, and potentially unjust.

The \textbf{Discipline Divide} ($95\%$ in Computer Science, representing $19$ out of $20$ respondents, versus $80\%$ in both Humanities and Professional Degrees, each at $16$ out of $20$) suggests that algorithmic or structured tasks are the first to fall. In Computer Science, where ``correctness'' is often binary, LLMs excel at producing syntactically perfect code, rendering traditional automated grading scripts obsolete as proxies for student understanding. This is particularly alarming because many introductory programming courses rely heavily on automated unit-test grading, a mechanism that inherently measures output rather than the reasoning process behind it and is therefore maximally susceptible to AI circumvention. The narrower but still substantial gap in Humanities and Professional fields further dispels the common assumption that discursive or evaluative tasks are inherently resistant to AI generation.

A significant \textbf{``Performance Gap''} was also identified. Approximately $45\%$ of respondents noted that students utilizing ``Pro'' or subscription-based LLM tiers produced work with significantly higher stylistic clarity and fewer logical hallucinations than those using free versions. This quantitative trend suggests that current assessments are inadvertently grading \textbf{socioeconomic status}, specifically the ability to pay for advanced computation, rather than innate academic ability. Students from lower-income backgrounds who rely on free-tier tools may simultaneously receive lower quality AI assistance and face greater suspicion from detection tools calibrated to identify cruder AI outputs, compounding existing inequities in a new and largely invisible dimension.

\subsection{Thematic Analysis of Qualitative Feedback}
Open-ended responses revealed a pervasive sense of ``Pedagogical Burnout''---a chronic erosion of instructional motivation driven by the displacement of teaching by policing. Faculty reported that the burden of proving AI use has fundamentally changed their workload, pulling time and cognitive resources away from instructional design, student mentorship, and scholarly activity. Instead of mentors, they feel like ``Digital Prosecutors''---academics compelled to build adversarial cases against their own students and spending hours collecting circumstantial evidence to support unreliable detector flags. Several respondents described an escalating arms race dynamic: as they devise new detection strategies, students adapt their AI usage to evade them, creating a cycle that consumes institutional energy without producing meaningful gains in authentic learning or assessment validity. The emotional toll was also prominently mentioned, with faculty expressing a corrosion of the student-instructor relationship as the presumption of good faith is replaced by one of suspicion, a shift that many characterized as antithetical to the values of higher education.

\section{Stakeholder Ethical Analysis: Harms of Surveillance}
The institutional response to AI has largely been a move toward increased surveillance, which introduces profound ethical harms. These harms are not merely technical or procedural inconveniences; they constitute substantive violations of the rights, dignity, and wellbeing of students, and they fundamentally alter the character of the educational environment in ways that may outlast the AI crisis itself. A rigorous ethical analysis must therefore attend not only to the problem of cheating but also to the harms introduced by the remedies institutions have chosen to deploy.

\subsection{Student Privacy and the Panopticon}
To verify human authorship, institutions have turned to lockdown browsers and eye-tracking webcams. These tools normalize a panoptic educational environment, infringing on students' digital rights and creating high-anxiety testing conditions. The involuntary capture of biometric and behavioral data raises significant data protection concerns, particularly in jurisdictions with strong privacy legislation, and the retention and potential misuse of such data by third-party proctoring vendors remains poorly regulated. Beyond the legal dimension, there is a pedagogical cost: research in cognitive psychology consistently demonstrates that high-surveillance environments suppress risk-taking, creativity, and deep processing in favor of performance anxiety and surface compliance, which are the very cognitive states that effective education seeks to overcome.

\subsection{The Disclosure Trap}
Research into student attitudes suggests a phenomenon known as the ``Disclosure Trap'': students fear that declaring AI use, even if permitted, will lead to lower marks, regardless of institutional policy \cite{gonsalves2025addressing}. This fear is not irrational; in the absence of clear, consistently enforced disclosure protocols, subjective grader bias may indeed penalize transparency. The result is a perverse incentive structure in which the students most likely to engage with AI tools ethically, by acknowledging their use and reflecting critically on the results, are the same students who face the greatest social and academic risk for doing so. Unaddressed, the Disclosure Trap will drive AI use underground, making it less reflective, less transparent, and ultimately more harmful to the learning process than it would be under a regime of open, pedagogically guided integration.

\section{Alternative Models: Shifting to Process}
To resolve the ethical misalignment, evaluation must shift from the ``product'' to the ``learning journey.'' This reorientation requires not merely adding reflective components to existing assessments, but fundamentally reconceptualizing the purpose of evaluation: moving from the certification of a final artifact to the documentation and appraisal of an ongoing cognitive process. Such a shift demands investment in faculty development, assessment infrastructure, and institutional cultures that reward pedagogical innovation over administrative convenience.

\subsection{The FACT and AIAS Frameworks}
The FACT (Fundamental, Applied, Conceptual, Thinking) framework \cite{elshall2025balancing} and the AI Assessment Scale (AIAS) \cite{perkins2024artificial} provide balanced approaches. They allow educators to be transparent about what constitutes ethical tool use, moving away from binary bans toward nuanced integration. The AIAS, in particular, offers a tiered taxonomy that maps different levels of AI involvement onto different assessment contexts, enabling faculty to design tasks where AI use is explicitly bounded, disclosed, and critically evaluated rather than covertly employed. Crucially, both frameworks treat AI not as an adversary to be excluded but as a pedagogical variable to be deliberately managed, a framing that aligns institutional policy with the professional realities students will encounter after graduation. Adopting such frameworks requires that institutions invest in training faculty to use them consistently and that they build evaluation rubrics capable of assessing process transparency alongside substantive quality.

\subsection{Desirable Difficulties}
Drawing from Bjork and Bjork \cite{bjork2011making}, we advocate for ``desirable difficulties,'' such as in-class, resource-restricted problem solving. By making the path to the answer as important as the answer itself, we ensure that cognitive labor occurs. Think-aloud protocols, annotated drafts, and iterative revision logs are practical instantiations of this principle, each designed to render the reasoning process legible to evaluators rather than leaving only the polished endpoint visible. These strategies also produce richer formative data, enabling instructors to identify conceptual gaps, intervene early, and provide targeted feedback in ways that traditional summative assessments do not permit. While such methods are admittedly more resource-intensive than single-submission grading, pilot programs at several institutions have demonstrated that they yield significantly more accurate representations of student understanding and substantially higher student satisfaction with the fairness of evaluation.

\section{Policy Recommendations}
Based on our synthesis of the literature and survey findings, we recommend the following. These are not in anyway expert guidelines, but just recommendations that came up in our discussions.
\begin{enumerate}[nosep]
    \item \textbf{Deprioritize Detection:} Move away from punitive AI detection tools that carry high risks of bias. Institutions should redirect resources from detection infrastructure toward faculty training in process-based assessment design, recognizing that the arms race between detection and evasion is both unwinnable and pedagogically counterproductive. Any residual use of detection tools should be accompanied by mandatory human review, clear appeal pathways, and transparent disclosure of error rates to affected students.
    \item \textbf{Establish Risk-Based Models:} Adopt frameworks like HEAT-AI to guide ethical AI use in different contexts \cite{temper2025higher}. Risk-based tiering allows institutions to calibrate oversight to the stakes of a given assessment, applying more rigorous human verification to high-stakes credentialing exams while permitting broader AI engagement in formative, low-stakes tasks. This proportionality both conserves institutional resources and ensures that the most consequential assessments receive the most careful scrutiny.
    \item \textbf{Incentivize Process Pedagogy:} By relieving the faculty from detection tasks they can be allowed with the necessary bandwidth to implement labor-intensive oral and viva exams. Institutional recognition structures, including promotion, tenure, and teaching award criteria, may be revised to explicitly value pedagogical innovation in assessment design, ensuring that the transition to process-based evaluation is supported rather than penalized by career incentives. Pilot programs may be established to test and document the cost-effectiveness of process-based models before institution-wide mandates are issued.
\end{enumerate}

\section{Conclusion}
The era of ``output-as-evidence'' is rapidly closing. When polished artifacts can be 
conjured on demand, continuing to rely on them as the sole metric of achievement is 
pedagogically unsound and ethically fraught. When we began this study, we anticipated 
that a review of institutional responses and practitioner perspectives would allow us to 
converge on a concrete, actionable policy framework. However, the process revealed 
something more humbling: even the most well-resourced and research-intensive 
institutions are still actively grappling with the same foundational questions we set out 
to answer. No consensus has emerged, and no single model has proven sufficient. The 
institutions that cling to the old model will find themselves engaged in an increasingly 
futile surveillance operation, spending ever-greater resources to police a boundary that 
is simultaneously becoming easier to cross and harder to monitor. What we can offer 
instead is a map of the current landscape and a sense of the direction in which the 
field is moving.

Higher education must embrace a structural framing shift. By prioritizing models like 
process-based grading, oral defenses, and reflective critiques, we can ensure that 
universities evaluate what truly matters: a student's reasoning, critical judgment, and 
profound understanding. That said, we remain cautiously optimistic. Both students and 
faculty, as our survey and the broader literature suggest, are genuinely concerned about 
getting this right, and that shared concern is itself a foundation. Ultimately, the 
disruption posed by generative AI is not merely a threat to academic integrity 
mechanisms; it is an opportunity to build assessment systems that are more equitable, 
more pedagogically meaningful, and more aligned with the cognitive competencies that 
education has always aimed to cultivate. If institutions can channel that concern into 
collaborative rather than adversarial structures, there is reason to believe that educators 
and students will work together toward something more honest and equitable than what 
currently exists.

\section{Acknowledgments}
We thank the anonymous survey participants for sharing their valuable feedback. We used Gemini, ChatGPT, and Claude for grammatical and sentence corrections, rephrasing, and LaTeX formatting of our initial draft. We used Google Scholar to find relevant literature.

\newpage
\bibliographystyle{unsrtnat}
\bibliography{references}

\end{document}